# ASTR2020: Training the Future Generation of Computational Researchers

State of the Profession Consideration: Computational Education

Lead Author: Gurtina Besla. (U. Arizona, gbesla@email.arizona.edu)

Co-Authors:

D. Huppenkothen (U. Washington, dhuppenk@uw.edu),
N. Lloyd-Ronning (LANL, lloyd-ronning@lanl.gov),
E. Schneider (Princeton, es26@astro.princeton.edu),
P. Behroozi (U. Arizona, behroozi@email.arizona.edu),
B. Burkhart (Rutgers/CCA, bburkhart@flatironinstitute.org),
C.K. Chan (U. Arizona, chanc@email.arizona.edu),
S.A. Jacobson (Northwestern, sethajacobson@earth.northwestern.edu),
S. Morrison (Missouri State, smorrison@psu.edu),
H. Nam (LANL, hnam@lanl.gov),
 S. Naoz (UCLA, snaoz@astro.ucla.edu),
A. Peter (OSU, peter.33@osu.edu),
E. Ramirez-Ruiz (UCSC, enrico@ucolick.org)

Endorsements from individuals:

Abrahams, E. (Berkeley), Bailin, J. (U Alabama), Blecha, L. (U. Florida), Bostroem, A. (UC Davis), Bogdanović, T. (Georgia Tech), Boylan-Kolchin, M. (UT Austin), Bellm, E. (U. Washington),  Bromm, V. (UT Austin), Brooks, A. (Rutgers), Bryan, G. L. (Columbia), Bullock, J. (UC Irvine), Chonacky, N. (Yale), Christensen, C. (Grinell), Cruz, K. (CUNY Hunter College), Donahue, M. (Michigan State), Eifler, T. (U. Arizona), Faucher-Giguère, C.-A. (Northwestern), Feigelson, E. (Penn State), Follette, K. (Amherst College), Ford, E. (Penn State), Gabriel, T. (ASU), Gnedin, O. (U. Michigan), Hayward, C. (CCA), Hernquist, L. (Harvard), Hilborn, R. (Associate Executive Director of AAPT), Ivezić, Z. (U. Washington), Johnston, K.V. (Columbia), Juneau, S. (NOAO), Jurić, M. (U. Washington), Kereš, D. (UCSD), Krause, E. (U. Arizona), Kravtsov, A. (U. Chicago), Lidz, A. (U. Penn), Madigan, A.-M. (Boulder), Markoff, S. (U. Amsterdam), Narayanan, D. (U. Florida), Norman, D. (NOAO), O'Shea, B. (Michigan State), Offner, S. (U. Amherst), Olsen, K. (NOAO), Paschalidis, V. (U. Arizona), Pinto, P. (U. Arizona), Ragozzine, D. (Brigham Young), Sales, L.V. (UCR), Siemiginowska, A. (CfA), Sipőcz, B. (U. Washington), Stone, J. (Princeton), Tchekhovskoy, A. (Northwestern), Tollerud, E. (STScI), Torrey, P. (U. Florida), Trac, H. (CMU), Turk, M. (U. Illinois), Valluri, M. (U. Michigan), Vogelsberger, M. (MIT), Walker, C. (NOAO), Walkowicz, L. (The Adler Planetarium), Weltman, A. (U. Cape Town),  Wetzel, A. (UC Davis), Wise, J. (Georgia Tech), Zaldarriaga, M. (IAS), Zhu, Z. (UNLV), Zingale, M. (Stony Brook)

LANL: A. Aykutalp, C. Biwer, P. Bradley, N. Butler-Craig, B. Dingus, E. Fogerty, C. Fryer, S. Gandolfi, M. Gilmer, F. Guo, J. Guzik, A. Hungerford, P. Kilian, O. Korobkin, J. Lippuner, H. Li, Z. Medin, J. Miller, M. Mumpower, P. Polko, M. Rassel, M. Taylor, B. Wiggins, R. Wollaeger

# ASTR2020: Training the Future Generation of Computational Researchers

## Executive Summary

**Motivation:** *The current disparity in computational knowledge and access across the nation is a critical hindrance to the diversity and therefore the success of the fields of Astronomy and Physics.* Thus, training the next generation of computational researchers is crucial. Preparing all students for careers in Astronomy, Physics, and industry will require universal computational and data science literacy, innovative education models, and equitable access to high performance computing (HPC) as well as elastic, specialized and other emerging computing facilities. The following recommendations are critical to the growth and retention of a new generation of computational researchers that reflect the demographics of the undergraduate population in Astronomy and Physics.

**Recommendation 1)** That the Decadal Survey strongly endorse the implementation of mandatory introductory computational science course requirements and the embedding of programming within core curricula, e.g., following recommendations from the American Association of Physics Teachers. These courses should incorporate four pillars of modern research: programming, best practices in software development, modern data analysis, and simulations. They should also use data sets, analysis methods, programming languages and codes commonly used by the profession. Specifically, we recommend that the Decadal endorse the following statement: ***The realities of modern Astronomy and Physics demand the institutionalization of computational literacy as a core competency, parallel in priority to basic math.***

**Recommendation 2)** That the Decadal Survey promote a community effort in partnership with, for example, *The Carpentries*, to create and maintain platforms for open-source computational curricula and instructor training. Open-source curricula will aid instructors across the country teach computational material at both introductory and advanced levels in colleges/universities, avoiding licensed course material and software. Explicit language supporting such an effort in calls for NSF Broader Impacts, NSF Postdoctoral Research Fellowships, NASA/DOE/NSF data management plans, coupled with dedicated or new funding lines, will be needed to incentivize programming that fulfills this recommendation. Furthermore, large science collaborations that receive operations funding from DOE/NSF/NASA should have a dedicated budget to create and publish curated tutorials based on their analysis software for use in both research and educational settings.

**Recommendation 3)** That the Decadal Survey endorse the development of HPC and advanced computational and data science training pathways for students and the funding of alternative education and training models beyond standard classrooms. This includes, but is not limited to: new REU programs, funded graduate and undergraduate summer/winter schools and hack sessions hosted at, e.g. national HPC facilities, institutions hosting large data sets (e.g. NOAO, STScI), data science institutes (e.g. eScience Institute at the University of Washington), and institutions with local HPC facilities, with the express purpose of offering methodological and computational training for students with *different levels of computational expertise*. Specifically, we recommend that NSF/NASA/DOE expand funding opportunities for programs aimed at increasing computational access and training for women and underrepresented minorities. As the computational pipeline grows (Rec. 1 and 2), a wider range of flexible funding streams and computational resources aimed at graduate students will be necessary to enable computationally intensive research, particularly for those students from underserved institutions.

**Recommendation 1: All institutions granting Astronomy, Astrophysics and Physics undergraduate and graduate degrees are strongly urged to require computational and data science classes at both introductory and advanced/applied levels**

*Undergraduate programs are growing and the relevant job markets require coding skills.* Over 2006-2017, the AIP reports that the number of institutions offering Astronomy degrees has increased by a factor of 2. Correspondingly, the total number of conferred Astronomy degrees has increased by a factor of 1.6 (Nicholson & Mulvey 2017). Upon graduation, Physics or Astronomy majors who manage to enter the private sector typically find employment in data science, computer systems, and engineering-based companies.[1] These professions require a range of computational skills including software development, data visualization, statistics, machine learning and working with large computational infrastructures. Astronomy and Physics undergraduate programs must prepare students for this reality and provide them with the appropriate practical and computational skill sets to be marketable in the private sector (Heron & McNeil 2016).

*Currently, the fields of Astronomy and Physics are driven by big questions that require computational solutions and massive data sets, making computational training imperative for careers in academia.* The "grand challenges" facing the field of Astronomy and Physics over the next 10 years will require innovative and advanced methodological and computational methods to solve. In particular, the fields of Astronomy and Physics are generating massive and complex data sets from large observational surveys (e.g., the Large Synoptic Survey Telescope, *LSST,* will produce 15 terabytes a night, Juric+2017), large-scale experiments (e.g., the Event Horizon Telescope produced 5 petabytes in the 2017 observation campaign, EHTC+2019; the LHC experiments produce over 50 petabytes of data per year;[2] LIGO generates terabytes a day[3]), and simulations (e.g., the IllustrisTNG Cosmological Simulations, 1.6 petabytes in total, Nelson+2019). The resulting petabyte-scale data sets require advanced computational methods and facilities to analyze (Zhang & Zhao 2015). Given these realities, it is imperative that the current generation of young astronomers and physicists are trained in computational methods. With coding skills in their toolkit early in their academic path, students can join and have greater impact in research groups earlier, solve meaningful problems in upper division and lab courses and obtain higher-paying technical summer internships.

*Dedicated introductory computational courses are not standard requirements within Physics and Astronomy core curricula, creating inequitable access to computational training pathways.* Given the above realities, it is increasingly clear that computational literacy is as essential as mathematics in modern Astronomy and Physics. However, few degrees have early computational requirements. For example, only 4 of the Pac-12 universities (which includes some of the largest public universities) have mandatory computational course requirements within freshman or sophomore years of Astronomy or Physics degrees. Furthermore, a recent national survey of 357 unique Physics departments find that "only 55% have a simple majority of their faculty reporting that they have some experience teaching computation to undergraduate physics students" (Caballero & Merner 2019). Inequity in early computational training can delay student entry into research and limits early exposure to computationally intensive research areas, such as theoretical research. As a result, students graduating from a subset of institutions have a significant advantage

---

[1] https://www.aps.org/careers/statistics/index.cfm
[2] https://home.cern/science/computing/storage
[3] https://www.ligo.caltech.edu/page/ligo-technology

in admission to and performance both during graduate school in Astronomy and Physics, and on the job market. *The current inequity in access to computational education is a major challenge to the inclusion and diversity of the field.*

*We recommend that all institutions conferring degrees in Astronomy, Astrophysics and Physics implement the following computational requirements, in a relevant computer language:*

A. <u>At least one required introductory course (3 credit hours) in programming in the freshman year of undergraduate programs in order to institutionalize programming as a core competency, parallel in priority to basic math.</u> Students enter Astronomy and Physics majors with a wide range of computational expertise, necessitating early access to training to level the playing field. An interdisciplinary approach, partnering with Computer Science or Engineering departments, can help meet a range of student needs and increase instructional capacity. However, in instances where, e.g., Computer Science classes are restricted to their majors, open-source introductory course material will be critical (Rec. 2). Adding new course requirements to Astronomy and Physics degrees, which typically have extensive requirements, may present a challenge for some institutions, as these additions must come at the expense of existing requirements or additional costs. Acknowledging this challenge, we stress that this recommendation is critically needed to meet the computational realities of modern careers in Astronomy and Physics, but we also caution graduate admissions committees to account for these institutional barriers in obtaining computing experience for potential admitted students.

B. <u>At least one upper-division class in computational methods, optimization, data science and statistics, with a focus on applications to Physics and Astronomy research.</u> With the advent of large surveys, computational datasets, and data intensive experiments, the methods of the modern astronomer and physicist have adapted. These relatively new skills must be taught alongside traditional methods such as astronomical observation and pen and paper calculations.

C. <u>Incorporation of programming into core undergraduate and graduate Physics and Astronomy courses (e.g. quantum mechanics, classical mechanics, thermodynamics, dynamics, cosmology, etc.).</u> With (A) in place, all upper division classes can assume students have a core competency in programming, enabling practical project-type exercises with a significant programming and data analysis component. As such, computing would not necessarily be the focus in these classes, but rather an inextricable component akin to mathematics. The American Association of Physics Teachers (AAPT) have made similar recommendations in 2011: "*The AAPT urges that every physics and astronomy department provide its majors and potential majors with appropriate instruction in computational physics*"[4], and have outlined specific learning outcomes and strategies for incorporating computation within core curricula[5]. We strongly echo these recommendations and emphasize working with the AAPT to make computing a normal component of Astronomy and Physics degrees.

---

[4] https://www.aapt.org/Resources/policy/Statement-on-Computational-Physics.cfm
[5] https://www.aapt.org/Resources/upload/AAPT_UCTF_CompPhysReport_final_B.pdf

**Recommendation 2: The Development and Maintenance of a Platform for Open-Source Computational Curricula and Programs for Instructor Training**

Open-source curricula have been extremely successful through the development of the *Software Carpentry* and *Data Carpentry* organizations. These programs, recently combined under "*The Carpentries,*"[6] train volunteer scientists to teach programming and data analysis methods in multi-day workshops around the world, and more crucially provide a global community working together at all levels of teaching including curriculum development, lesson infrastructure and maintenance, development of assessment, and instructor training. We recommend extending instructor training programs and curricula that exist as part of, e.g. *The Carpentries,* in order to build such a community at a smaller scale for computational astrophysics.

*A per-instructor academic teaching model impedes the creation of computational training pathways.* Traditional academic teaching is often based on a per-instructor model, where the faculty member responsible for teaching the class is also solely responsible for the development and improvement of that class. The lack of instructor knowledge is found to be one of the major difficulties associated with integrating computation into undergraduate physics (Leary+2018). Well-designed curricula on an open platform will lower the barrier to entry for instructors without a strong background in computational methods to teach needed classes (Rec. 1). However, barriers will likely remain for upper division/graduate classes, necessitating dedicated instructor training pathways, such as the *Software Carpentry Instructor Training Program.*[7]

*An open-source platform will increase efficiency.* In a per-instructor academic teaching model, every lecturer reinvents the wheel by designing their own curriculum from scratch, leading to duplication of effort. Computational and data science courses are more hands-on than traditional courses and typically require more development time for modules by comparison. With an open-source platform, initial curricula do not have to be polished, but can serve as a blueprint that allows the community to improve upon and personalize. As such, the burden of developing lectures need not fall upon a single person. Instead, researchers can add single components (such as a single lecture, tutorial or an exercise).

*Keeping up with the pace of platform development/code changes*. Computational methods and technologies change at a much more rapid pace than changes in our knowledge of fundamental physics. For example, best practices for software development published two years ago may already be outdated today. It is therefore critical that curricula in Computational Astrophysics, where results often depend crucially on programming languages, platforms and software packages, keep pace with these changes. For example, while Python is currently a favored language, there is no guarantee that this will be the case in ten years. Even Python has experienced rapidly evolving features and libraries that require frequent updates of software. While this constant change may be daunting, it is indicative of the broader utility of a computational education and the vast industry accessible with such education. With an open-source platform, multiple instructors can provide incremental updates, allowing adaptations to developments in computation on shorter timescales.

*Open-source development of curricula and lessons allow contributions from a diverse pool of educators,* allowing for collaboration across disciplines, nationalities, institution sizes, and career stages. This allows researchers often excluded from formal teaching (e.g. early-career researchers or researchers at non-university institutions) to offer their expertise. In addition, the open-source format allows for easy sharing and incorporation of curricula and lessons taught during more informal teaching arrangements such as summer schools and workshops (Rec. 3). Paid course and lesson maintainers provide quality control for each semester-level course, as well as individual lessons, enabling a wide range of researchers to gain experience in curriculum development. This structure, coupled with the community-driven improvement model of Open-Source, will allow for continuous improvement of existing lesson, as well as quick adaptation to computational novelties.

---

[6] https://docs.carpentries.org/
[7] https://carpentries.github.io/instructor-training/

***Maximizing the impact of state-of-the-art theoretical simulations, astrophysical data sets and codes***. The *IllustrisTNG* public data release (Nelson+2019), the *Catalog for Astrophysical Turbulence Simulations* (CATS)[8] and *MESA-Web*[9] are fantastic examples of how theorists are making simulations available to the community. Similarly, a number of astronomical data sets are publicly available to researchers for a broad range of science cases, including *Gaia, Kepler, TESS* and *LSST*. However, the average graduate student does not have the training to properly analyze and interpret these data sets. <u>The impact of these releases to the broader community would be significantly augmented if training sets, curricula, and individual lessons were released using the simulations or data.</u> Without the development of such educational tools, the analysis and impact of new data will stay within the existing user base or scientific community (see Rec. 3). This is similarly true for the usage of standard codes, like *Athena, Cloudy, Einstein, Enzo, HARM, Gadget, Pencil, Toolkit,* etc., and newer codes like *Cholla, GAMER2, Trident*, etc. <u>With an open-source platform, training sets need not be perfect or created by the team that developed and released the simulation or code.</u> Impressive examples of this kind of community service already exist, e.g., the *pyro* code[10] (Harpole, Zingale+2019) and *AstroML*[11] (VanderPlas, Connolly+2012).

## *We recommend:*

A. <u>A community effort, building on the experiences and materials available through *The Carpentries* organizations as a foundation to create an open-source platform for curriculum development and instructor training.</u> This will allow for curricula to be continuously updated to keep lecture material and the tools and methods being taught relevant to students' future careers, both within academia and industry. Some examples of open-source platforms for curricula/training are already available: 1) AAPT, Partnership for Integration of Computation into Undergraduate Physics (*PICUP*)[12]; 2) on GitHub in the *Open Astro Bookshelf*[13]; and 3) *HPC Carpentry*[14] hosts advanced computing curricula. Maintaining semester-long courses, however, will require institutional support and broad oversight committee.

B. <u>Explicit language in calls for NSF Broader Impacts, NSF Postdoctoral and DOE Graduate Research Fellowships, DOE/NASA/NSF data management plans and dedicated grant programs to incentivize programs that fulfill the above recommendations.</u> For example, the creation and maintenance of platforms for open-source computational curricula designed to provide all students with tools to analyze the funded data products should be explicitly mentioned as an example of a possible 'Broader Impact[15].' Specifically, large science collaborations that receive operations funding from DOE/NSF/NASA should be required to publish research and educational tutorials based on their main analysis software and provide avenues for instructor training.

---

[8] http://www.mhdturbulence.com/CATS.html
[9] http://mesa-web.asu.edu/
[10] https://github.com/python-hydro/pyro2
[11] http://www.astroml.org
[12] https://www.compadre.org/PICUP/
[13] https://github.com/Open-Astrophysics-Bookshelf
[14] https://hpc-carpentry.github.io/

**Recommendation 3: Development of computational training pathways for students through increased access to national HPC facilities and increased funding for alternative education models beyond standard classrooms.**

***Formalized, universal training pathways for the usage of state-of-the-art HPC facilities and modern data analysis infrastructures do not currently exist.*** Training pathways for undergraduates and graduate students that commonly exist in observational Astronomy and experimental Physics ultimately equip students to use state-of-the-art observational/experimental facilities. The absence of comparable HPC and methodological training pathways reflects outdated perceptions of the basic research skills needed to succeed in these fields, formalized in an era prior to the current reality of massive data sets. Instead, students are forced to "pick up" advanced computational skills informally, largely through research experiences, often resulting in only a cursory knowledge of HPC, cloud computing, statistics and machine learning, and a limited eligible user pool for national HPC.

***Graduate student access to national HPC facilities is limited.*** Most institutions do not have local HPC facilities or instructors trained in their usage. National HPC facilities are thus an important pathway for computational research at smaller institutions that may not have the necessary institutional resources. However, graduate students cannot be PIs on proposals for the majority of HPC facilities (e.g. LANL and DOE facilities). Only NSF graduate student fellows can apply to XSEDE startup allocations, but even they must have a faculty PI if they wish to request larger allocations. There exists an acute imbalance in access to state-of-the-art computing for graduate students across the country.

***PhD. theses in Computational Astrophysics increasingly rely on HPC and advanced computational methods, both for generating and analyzing massive simulations and observational data sets.*** Most areas of Astronomy research critically require advanced computational methods (such as modern statistics and Machine Learning) and the usage of HPC facilities. Indeed, the usage of HPC facilities and advanced computing techniques are common among the list of research conducted by awardees of NASA Postdoctoral Fellows over the past 10 years[16]. The lack of training and equitable access to HPC facilities and advanced computational and statistical methods are currently significant barriers to pursuing computational research at the graduate level. The resulting disparity in computational knowledge across the nation is a critical hindrance to the diversity of the field.

***Informal learning is a pathway to universal computational literacy.*** There are, however, models of education beyond the traditional classroom format. *Summer/Winter schools* have long been known as an effective format to disseminate recent results through the community and train students in advanced methods (e.g. the *LSST Data Science Fellowship Program*[17], the *Penn State Summer School in Statistics for Astronomers*[18], the *La Serena School in Data Science*[19]). In recent years, new formats have emerged, including *Massively Open Online Classes* (MOOCs) that allow dissemination of information to a wide audience; *hack weeks*, which combine theoretical learning more akin to traditional classes, and summer schools with practical project work (Huppenkothen+2018), such as the successful OLCF-based GPU hackathons[20] and *Astro Hack Weeks*[21]; and student competitions that engage undergraduates from academic institutions with less access to computational resources, such as the *Student Cluster Competition*[22]. In particular, the *Student Cluster Competition* serves as a microcosm of the real-world HPC Community and has been shown to both engage and educate students who have not had HPC exposure.

---

[15] See also ASTRO2020 White Paper by Norman et al "*Tying Research Funding to Progress on Research Inclusion*"
[16] http://www.stsci.edu/stsci-research/fellowships/nasa-hubble-fellowship-program/meet-the-fellows

***Existing computational summer schools are not meeting student or instructor needs.*** Computational summer schools generally have a high impact on students, providing them with essential computational skills for their careers (e.g. IHPCSS[23]). At Los Alamos National Laboratory (LANL), for example, the retention rate of students attending HPC summer schools (i.e. students who return to work at LANL in fields involving HPC) is ~50%, compared to ~20% with other summer programs (H. Nam, private communication 2019). However, existing programs teaching data-intensive skills like the *LSST Data Science Fellowship Program* are massively oversubscribed. In addition, there is typically a high bar of computational knowledge required for acceptance into existing HPC summer schools (e.g. ATPESC, LANL), which explicitly require significant prior experience with HPC techniques (e.g. MPI). <u>There is no formal process or pipeline for reaching students with less preparation or less previous access to resources.</u> Furthermore, these schools accept a small fraction of applicants, and many existing programs are not well funded or valued, as evidenced by the recent cancellation of NSF funding for the *Blue Water's Student Internship*[24] and discontinuation of funding for UC-HiPACC.[25] Finally, there exist only a handful of programs that train mentors in HPC or data science (e.g. the NSF-sponsored *Center for Parallel and Distributed Computing Curriculum Development*[26], *The Carpentries*), some of which reach only a small number of faculty members (~20), again highlighting the need for a widespread, open-source curriculum easily accessible by mentors and students alike (Rec. 2).

***We recommend*** *the following informal training paths, best practices and funding structures to level the playing field for all students pursuing careers in computational research:*

A) <u>The expansion of student training opportunities in data science and on national HPC facilities,</u> including introductory versions of existing HPC summer schools with a more flexible bar for entry, increased funding and expansion of existing student training programs in data-intensive research, student competitions/programs that engage students at institutions without local HPC resources, and reinstatement of programs like the Blue Waters Summer Internship. We further recommend that NSF/NASA/DOE increase funding opportunities for programs specifically aimed at increasing computational access/training for women and underrepresented minorities, such as the *Advancing Theoretical Astrophysics*[27] summer school. Offered trainings should serve as certification for student access to specific facilities.

---

[17] https://astrodatascience.org
[18] https://astrostatistics.psu.edu/su19/
[19] http://www.aura-o.aura-astronomy.org/winter_school/
[20] https://www.olcf.ornl.gov/for-users/training/gpu-hackathons/
[21] http://astrohackweek.org/2019/
[22] http://studentclustercompetition.us/Education/index.html
[23] http://www.ihpcss.org/
[24] https://bluewaters.ncsa.illinois.edu/internships
[25] http://hipacc.ucsc.edu/SummerSchool.html
[26] https://grid.cs.gsu.edu/~tcpp/curriculum/?q=node/21620

B) <u>That national HPC facilities reserve a dedicated amount of time on these facilities for startup allocations available for graduate student PIs, particularly from underserved institutions.</u> Graduate students are routinely PIs on proposals for state-of-the-art observational facilities. With the establishment of HPC training pathways, such as through (A), and the existence of support staff for national HPC facilities, there is no clear reason why graduate students should not be PIs of their own allocations. It will be impossible to grow a diverse user base of national computational facilities if junior users are required to apply with an experienced faculty advisor, as this limits the user base primarily to a subset of R1 institutions. <u>We further recommend that faculty be allowed to support multiple proposals led by graduate students</u>. Students should not be in competition with their peers for support from the same PI. We stress the need for the strong growth of open HPC infrastructure to support a growing user base.[28]

C) <u>That funding agencies open up a wider range of flexible funding streams aimed at graduate students for computationally intensive research.</u> While HPC resources are finite, new cloud-based architectures allow computation at scale. At the same time, new hardware dedicated to certain applications (e.g. Google's Tensor Processing Units) undergoes rapid cycles of development, and university computing architectures may not be able to respond to hardware, software and infrastructure needs on timescales required by modern research. While some viable routes exist for university/industry partnerships for graduate students (e.g. AWS Cloud Credits for Research[29] and NVidia's GPU Grants for Researchers[30]), public funding streams should enable early-career researchers to work with state-of-the-art computational methods, irrespective of their home institution. Potential models for such an infrastructure include the Open Science Data Cloud[31] and SciServer.[32]

D) <u>That language be added to calls for NSF REU site proposals that explicitly focus on data science, computational training and HPC access.</u> Such programs could be hosted at both national HPC facilities and institutions hosting massive data sets and/or with local HPC capabilities/in-house expertise. These undergraduate computational research experiences can also serve as training certification programs for students that can bolster applications for e.g. summer computational schools or allocations on HPC facilities as graduate students (see B).

E) <u>That calls for NSF Broader Impacts include dedicated budgets and language recognizing non-traditional computational training pathways as viable forums to improve computational literacy, as they directly address disparities in access to computational training.</u> Examples include combinations of existing and newly developed MOOCs (e.g. using the open-source teaching platform suggested in Rec. 2), project-based summer/winter schools and hack weeks.

---

[27] https://collectiveastronomy.github.io/advancingtheoastro/
[28] https://www.nap.edu/read/18972/chapter/1
[29] https://aws.amazon.com/research-credits/
[30] https://developer.nvidia.com/academic_gpu_seeding
[31] https://www.opensciencedatacloud.org
[32] http://www.sciserver.org/

F) <u>An updated model for summer/winter schools and hack weeks that moves away from chalkboard-based lectures towards modern teaching practices,</u> including flipped classrooms and project-based learning. Blueprints for such events, where lectures are actively combined with student-centered project work, exist both in the Astronomy community (e.g. LSST Data Science Fellowship Program[33], La Serena School for Data Science[34]) and in other fields (e.g. the Woods Hole Computational Neuroscience Summer course[35]). In addition, we recommend exploring peer learning through participant-driven events such as hack weeks, encouraging collaborations and networking. <u>Ultimately, the field must acknowledge completed programs of this type when considering candidates for research positions in Astronomy and Physics.</u>

G) <u>We call upon the community to transparently implement best practices for cohort selection for the above programs and provide funding for participants from underrepresented groups, small institutions and countries with limited travel funding</u>. In-person events are almost always space-limited, such that only a fraction of applicants can participate, and therefore may exacerbate existing structural inequalities in the field. In addition, <u>enabling remote participation for parts of or all of an in-person event</u> e.g. through live-streamed and recorded lectures, and chat apps such as Slack will drastically increase the reach of an event.

**Final Remarks**

The 2018 report from the Committee on STEM Education of the National Science & Technology Council explicitly states that in order to build a strong foundation for STEM Literacy we must ensure that "every American has the opportunity to master basic STEM concepts, including computational thinking," meaning "the ability to solve complex problems with data and computational methods"[36].

Over the next ten years, computation will play an increasing role in all fields of Theoretical Physics and Astrophysics, Experimental Physics and Survey Science. In order to harness the data revolution and address the biggest problems in the sciences, all elements of the scientific method must become scalable: hypotheses must be formulated into computation codes; experiments must involve automation; refinement must be assisted by artificial intelligence. "Computational thinking" is thus arguably *the* critical skill for student success in Astronomy and Physics. The recommendations in this document are designed to build computational training pathways for students that are both equitable and in keeping with the state-of-the-art.

Furthermore, the recommendations made in this document are central to the realization of five of NSF's 10 Big Ideas in the fields of Astronomy and Physics:

1) *Harnessing the Data Revolution[37],* which has the stated goal of developing "a cohesive, federated, national-scale approach to research data infrastructure, and the development of a 21st-century data-capable workforce."

---

[33] https://www.lsstcorporation.org/fellowship_program
[34] http://www.aura-o.aura-astronomy.org/winter_school/
[35] https://www.mbl.edu/mcn/
[36] *"Charting a course for success: America's Strategy for STEM Education"* Dec 2018 https://www.whitehouse.gov/wp-content/uploads/2018/12/STEM-Education-Strategic-Plan-2018.pdf
[37] https://www.nsf.gov/news/special_reports/big_ideas/harnessing.jsp

2) *NSF INCLUDES*[38], which seeks to "transform education and career pathways to help broaden participation in science and engineering."

3) *Growing Convergence Research*[39], which acknowledges that the grand challenges of today, such as exploring the universe at all scales, will not be solved by one discipline alone, but rather "requires convergence: the merging of ideas, approaches and technologies from widely diverse fields of knowledge." For example: capturing the first image of black hole from the *EHT* requires astronomical data processing and computational image reconstruction; and developing the transient classification and prediction algorithms for the *LSST* requires astrophysical models of different transient events and knowledge in machine learning algorithms. Such interdisciplinary research can only be done efficiently if both software and training materials are open source.

4) *Windows to the Universe*, which seeks to combine knowledge through "diverse windows - electromagnetic waves, high-energy particles and gravitational waves" in order to probe and understand the universe. Executing this vision will require a computationally literate workforce that can develop and utilize new analysis capabilities (neural networks, machine learning, etc.) to correlate the massive amounts of data being generated in each of these windows. Of particular relevance to Astronomy and Physics is the amount of data expected to be generated in the era of LISA and the daunting computational task of relating these signals to EM sources.

5) *Quantum Leap*, which aims to exploit quantum mechanics to develop "next-generation technologies for sensing, computing, modeling, and communicating." It is imperative to build a computationally literate population in Physics and Astronomy to realize this goal.

Ultimately the recommendations laid out in this paper will promote the growth and development of a user base for large astronomical data sets and national HPC facilities, creating a new generation of computational researchers that reflect the demographics of the undergraduate population in Astronomy and Physics. Specifically, these recommendations will enable:

- The creation of a socio-economically diverse, computationally literate workforce in Astronomy and Physics through early exposure to computational research and the institutionalization of computational literacy as a core competency, parallel in priority to basic math, within normal undergraduate/graduate education. Lowering barriers to entry and increasing both training and access to computational resources will create new pathways for a diverse student body to pursue computational research.

- The empowerment of educators to develop/document/update data analysis software in keeping with the state-of-the-art and fostering of a growing community that recognizes software development as a needed skill set for the health and growth of Astronomy and Physics in the era of massive data sets.

- The expanded usage/impact/reach of massive simulations/data, advanced codes and national computational facilities through increased computational literacy. These structures are often funded by large governmental grants, but, owing to the current sporadic nature of computational education, these structures impact a minority of students in Astronomy and Physics.

---

[38] https://www.nsf.gov/news/special_reports/big_ideas/includes.jsp
[39] https://www.nsf.gov/news/special_reports/big_ideas/convergent.jsp

**Acknowledgements:** The authors thank Sera Markoff and Enrico Ramirez-Ruiz for their organization of the "*Special Theoretical Astrophysics Workshop*" in Copenhagen, Denmark from July 9-13th 2018. This conference elevated the voices of minorities in computational astrophysics and the discussions that ensued inspired the creation of this paper.

**Many of the views expressed in this ASTRO2020 white paper are also echoed in that of:**

>**Norman+2019 "*The Growing importance of a Tech Savvy Astronomy and Astrophysics Workforce*"**
>
>**Desai+2019 "*A Science Platform Network to Facilitate Astrophysics in the 2020s*"**
>
>**and past white papers, such as Zingale+2016, "*The Importance of Computation in Astronomy Education*".**

*The views expressed in this white paper are not necessarily the views of the AAS, its Board, or its membership.*